\documentclass[useAMS,usenatbib]{mn2e}
\usepackage{graphics,epsfig,lscape,txfonts,color}
\usepackage[english]{babel}
\def\deg{\hbox{$^\circ$}}
\newcommand{\mr}{\mathrm}

\newcommand{\ergcm}[1]{$\times10^{#1}$~\hbox{erg~cm$^{-2}$~s$^{-1}$}}

\newcommand{\ergs}[1]{$\times10^{#1}$~\hbox{erg~s$^{-1}$}}

\def\ie{i.\,e.}                                      % i.e. (kursiv) \ie
\def\eg{e.\,g.}                                      % e.g. (kursiv) \eg

\def\xmm{\textit{XMM-Newton}}

\def\gx339{GX\,339-4}
\def\h1743{H\,1743-322}

\title[Energy dependent variability of \h1743 in 2014]{Using \xmm\ to study the energy dependent variability of \h1743\ during its 2014 outburst\thanks{Based on observations obtained with \xmm, an ESA science mission with instruments and contributions directly funded by ESA Member States and NASA.}}
\author[H. Stiele, W. Yu]{H. Stiele$^{1,2}$\thanks{E-mail:
hstiele@mx.nthu.edu.tw}, W. Yu$^{2}$ \\
$^{1}$Institute of Astronomy and Department of Physics, National Tsing Hua University, No.~101 Sect.~2 Kuang-Fu Road, Hsinchu, 30013, Taiwan\\
$^{2}$Shanghai Astronomical Observatory and Center for Galaxy and Cosmology, 80 Nandan Road, Shanghai, 200030, China}% \\
\begin{document}

\date{Accepted 2016 April 7. Received 2016 April 7; in original form 2015 May 5}
%\date{2016 April 7}

\pagerange{\pageref{firstpage}--\pageref{lastpage}} \pubyear{2016}

\maketitle

\label{firstpage}

\begin{abstract}
Black hole transients during bright outbursts show distinct changes of their spectral and variability properties as they evolve during an outburst, that are interpreted as evidence for changes in the accretion flow and X-ray emitting regions.
We obtained an anticipated XMM-Newton ToO observation of \h1743\ during its outburst in September 2014. Based on data of eight outbursts observed in the last 10 years we expected to catch the start of the hard-to-soft state transition. The fact that neither the general shape of the observed power density spectrum nor the characteristic frequency show an energy dependence implies that the source still stays in the low-hard state at the time of our observation near outburst peak. The spectral properties agree with the source being in the low-hard state and a \textit{Swift}/XRT monitoring of the outburst reveals that \h1743\ stays in the low-hard state during the entire outburst (a.\,k.\,a.~`failed outburst'). 
We derive the averaged QPO waveform and obtain phase-resolved spectra. Comparing the phase-resolved spectra to the phase averaged energy spectrum reveals spectral pivoting.
We compare variability on long and short time scales using covariance spectra and find that the covariance ratio does not show an increase towards lower energies as has been found in other black hole X-ray binaries. There are two possible explanations: either the absence of additional disc variability on longer time scales is related to the rather high inclination of \h1743\ compared to other black hole X-ray binaries or it is the reason why we observe \h1743\ during a failed outburst. More data on failed outbursts and on high-inclination sources will be needed to investigate these two possibilities further.
\end{abstract}


\section{Introduction}
\h1743, located at a distance of 8.5$\pm$0.8 kpc \citep{2012ApJ...745L...7S}, was discovered by the \textit{Ariel V} satellite during a bright outburst in 1977 \citep{1977IAUC.3099S...1K}. Observations with the \textit{HEAO-I} satellite allowed to determine the position more accurately and ruled out an association with 4U 1755--388 \citep{1977IAUC.3106....4K}. 
It belongs to the few sources where X-ray jets have been imaged \citep{2005ApJ...632..504C}.  
\h1743\ has very frequent outbursts, unlike most of the other black hole transients (BHTs). It showed its brightest outburst in 2003 ($\sim$100 cts/s in RXTE/ASM) that was studied by RXTE and INTEGRAL \citep{2003A&A...411L.421P,2005ApJ...623..383H,2005ApJ...622..503C,2005ApJ...629.1008J,2009ApJ...698.1398M}, followed by outbursts in September 2004, September 2005, and January 2008 \citep{2008ATel.1378....1K}. After the source showed a ``failed'' outburst in October 2008 \citep{2009MNRAS.398.1194C} it showed another (``full'') outburst in July 2009 \citep{2010MNRAS.408.1796M}.  Three additional outbursts have been observed in December 2009/ January 2010, August 2010 and April 2011 \citep{2013MNRAS.431.2285Z}. Further outbursts have been reported in December 2011/January 2012  \citep{2012ATel.3842....1N}, September 2012 \citep{2012ATel.4419....1S,2014ApJ...789..100S} and August 2013 \citep{2013ATel.5241....1N}. Most outbursts reached around 20 cts/s in RXTE/ASM 1.5 -- 12 keV range, with the ``failed'' outburst in 2008 being the faintest one with $\sim$10 cts/s in RXTE/ASM \citep[see Fig.~1 in][for a longterm RXTE/ASM light curve]{2013MNRAS.431.2285Z}. Based on the X-ray dipping behaviour observed in one observation it is believed that the inclination angle of its accretion disc is relatively high ($>$70$^{\circ}$) to our line of sight \citep{2005ApJ...623..383H}. A systematic study of the effect of orbital inclination on the shape of the HID and the accretion disc temperature supports a high inclination angle for \h1743 \citep{2013MNRAS.432.1330M}. 

Regarding the outburst evolution of BHTs there are two distinct patterns. Most BHTs evolve from the low hard state (LHS) through the hard and soft intermediate states (HIMS, SIMS) into the high soft state (HSS), before they return at a substantial lower luminosity passing again the intermediate states \citep{2011MNRAS.418.1746S} back to the LHS. Some BHTs stay in the LHS during the entire outburst \citep[see][for a list of BHTs that show LHS-only outbursts]{2009MNRAS.398.1194C}. Individual states have distinct timing properties. Power density spectra (PDS) obtained in the LHS are dominated by band-limited noise (BLN) and can show quasi-periodic oscillations (QPOs). When the BHT goes from the LHS to the HIMS the PDS obtained in the energy range covered by the RXTE/PCA detector (\ie\ $>$2 keV) show an increasing amplitude and centroid frequency of the (type-C) QPO \citep[see][for a classification of the different QPO types]{1999ApJ...526L..33W,2005ApJ...629..403C,2011BASI...39..409B}. Investigations of PDS in different energy bands, including softer energies, revealed the existence of two distinct power spectral shapes at soft and hard X-rays at the onset of the HIMS \citep{2013ApJ...770..135Y,2014MNRAS.441.1177S}. 

In 2008 \h1743\ showed an outburst in which it went to the HIMS, but then returned to the LHS without ever reaching the soft states. Compared to previous outbursts, where the source went to the HSS, the inner disc temperature did not vary substantially during outburst evolution and remained rather high for a HIMS ($T_{\mr{in}}\approx0.8$keV), while the inner disc radius decreased and the flux of the blackbody component increased \citep{2009MNRAS.398.1194C}. \citet{2009MNRAS.398.1194C} supposed that the returning to the LHS without entering into the soft states was connected to a premature decrease of the mass accretion rate. This kind of outburst that does not make it to the soft state was dubbed ``failed" outburst by \citet{2009MNRAS.398.1194C}. This naming is somewhat misleading since an outburst has taken place and as no outburst is bound to reach the soft state.

\subsection{The 2014 outburst}
\begin{figure}
\resizebox{\hsize}{!}{\includegraphics[clip,angle=0]{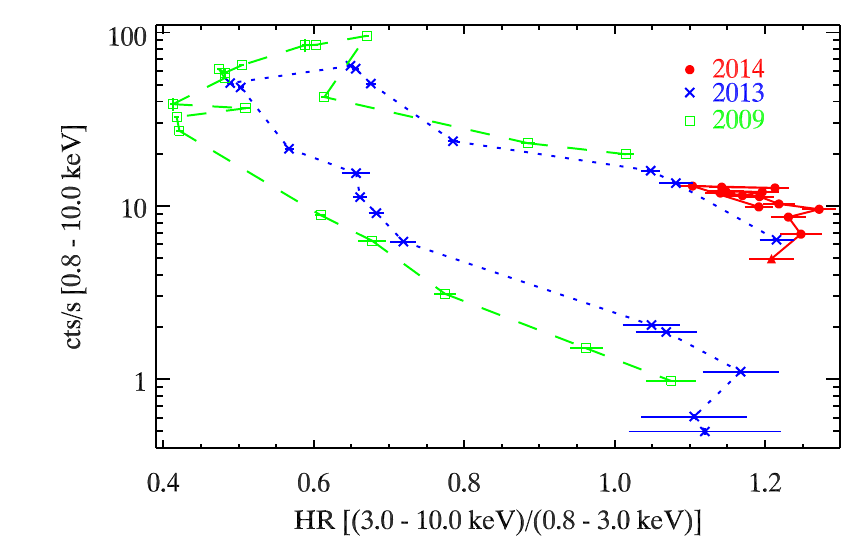}}
\caption{Hardness intensity diagram derived using \textit{Swift}/XRT window timing mode data, obtained between September 16 (red triangle) and October 10. For comparison the HIDs of the 2013 (blue X) and 2009 (green squares) outbursts are shown.}
\label{Fig:HID}
\end{figure}

To get an impression of the behaviour of \h1743\ during its outburst in 2014, we derived a hardness intensity diagram \citep[HID;][]{2001ApJS..132..377H,2005A&A...440..207B,2005Ap&SS.300..107H,2006MNRAS.370..837G,2006csxs.book..157M,2009MNRAS.396.1370F,2010LNP...794...53B} using \textit{Swift}/XRT \citep[The X-ray Telescope;][]{2000SPIE.4140...64B,2000SPIE.4140...87H} data. \textit{Swift}/XRT window timing mode data were taken between September 16 and October 10. The hardness ratio was obtained by dividing the count rate in the 3.0 -- 10.0 keV band by the count rate in the 0.8 -- 3.0 keV band. From the HID (Fig.~\ref{Fig:HID}) it is obvious that the source showed very little spectral evolution and that its spectrum remained rather hard during the entire outburst. In the observation that is softest the source still shows a fractional rms value above 30 per cent. Such high variability is only seen when a BHT is in the LHS, while in the soft states rms values below 5 per cent are observed \citep[see \eg\ ][]{2011MNRAS.410..679M}. The energy spectrum obtained during this observation can be described well ($\chi^2$/dof=949.1/817) by an absorbed power law with a photon index of $1.73^{+0.06}_{-0.05}$. The spectral and timing properties indicate that the source remained hard and that \h1743\ did not reach the soft states during its 2014 outburst.

\section[]{\xmm\ observation and data analysis}
\label{Sec:obs}
In this paper, we present a comprehensive spectral and variability study of \h1743\ based on an \xmm\ \citep{2001A&A...365L...1J} observation obtained on September 24th 2014 (PI: Stiele; exposure time 51.436~ks).  This observation was taken during outburst rise, about ten days after the detection of renewed activity of \h1743\ by INTEGRAL \citep{2014ATel.6474....1D}. A series of \textit{Swift} ToO observations were proposed to trace the low frequency QPO and spectral evolution in \h1743, which helped us to trigger an observation with \xmm\ of its brightest hard state usually expected just before a hard-to-soft transition. \h1743\ was observed with the EPIC/pn camera \citep{2001A&A...365L..18S} in timing mode. We filtered and extracted the pn event file, using standard SAS (version 14.0.0) tools, paying particular attention to extract the list of photons not randomized in time. As the SAS task \texttt{epatplot} showed no significant deviations of the observed pattern distribution from the theoretical prediction in the 0.8 -- 8 keV range, the observation was not affected by pile-up and we extracted source photons from a 15 column wide strip centred on the column with the highest count rate (31$\le$RAWX$\le$45). We selected single and double events (PATTERN$\le$4) for our study.
 
\subsection{Timing analysis}
We extracted light curves and produced power density spectra (PDS) in different energy bands (individual bands are specified in Sect.~\ref{Sec:res})  
for the longest continuous exposure available (a segment of 31.8~ks). After verifying that the noise level at frequencies above 30~Hz is consistent with the one expected for Poissonian noise \citep{1995ApJ...449..930Z}, we subtracted the contribution due to Poissonian noise, normalised the PDS according to \citet{1983ApJ...272..256L} and converted to square fractional rms \citep{1990A&A...227L..33B}. The PDS were fitted with models composed of zero-centered Lorentzians for BLN components, and Lorentzians for QPOs following \citet{2002ApJ...572..392B}. 

Furthermore, we derived covariance spectra following the approach described in \citet{2009MNRAS.397..666W}. 
As reference band we used the energy range between 1 and 4 keV, taking care to exclude energies from the reference band that are in the channel of interest. 

\subsection{Spectral analysis} 
\label{SubSec:XMMSpec}
We also extracted the averaged energy spectrum. As for the timing study we used the EPIC/pn data and selected source photons from a 15 column wide strip in RAWX centred on the column with the highest count rate, including single and double events. No energy selection has been applied here. Background spectra have been extracted from columns 3 to 5. Using standard SAS tools (\texttt{rmfgen} and \texttt{arfgen}) and the latest calibration files the corresponding redistribution matrix and ancillary response file have been determined.

Since energy spectra obtained from EPIC/pn fast-readout mode data are known to be affected by gain shift due to Charge-transfer inefficiency \citep[see][]{2014A&A...571A..76D} and show excess emission below $\sim$1 keV \citep[see \eg\ ][]{2006A&A...448..677M}, we also produced calibrated RGS \citep[Reflection Grating Spectrometers;][]{2001A&A...365L...7D} event lists, energy spectra, and response matrices using the SAS task \texttt{rgsproc}. Due to the presence of this soft excess emission we limited our spectral studies to energies above 0.8 keV, the limit suggested in the EPIC calibration status report\footnote{version 3.7, edited by M.~J.~S.~Smith (ESAC) on behalf of the EPIC Consortium; available at http://xmm2.esac.esa.int/docs/documents/CAL-TN-0018.pdf}. We included lower energies in the case of spectral ratios, as the division should remove the contribution of the excess emission, assuming that it is constant on the time scales considered.

\section[]{Results}
\label{Sec:res}

\begin{figure}
\resizebox{\hsize}{!}{\includegraphics[clip,angle=0]{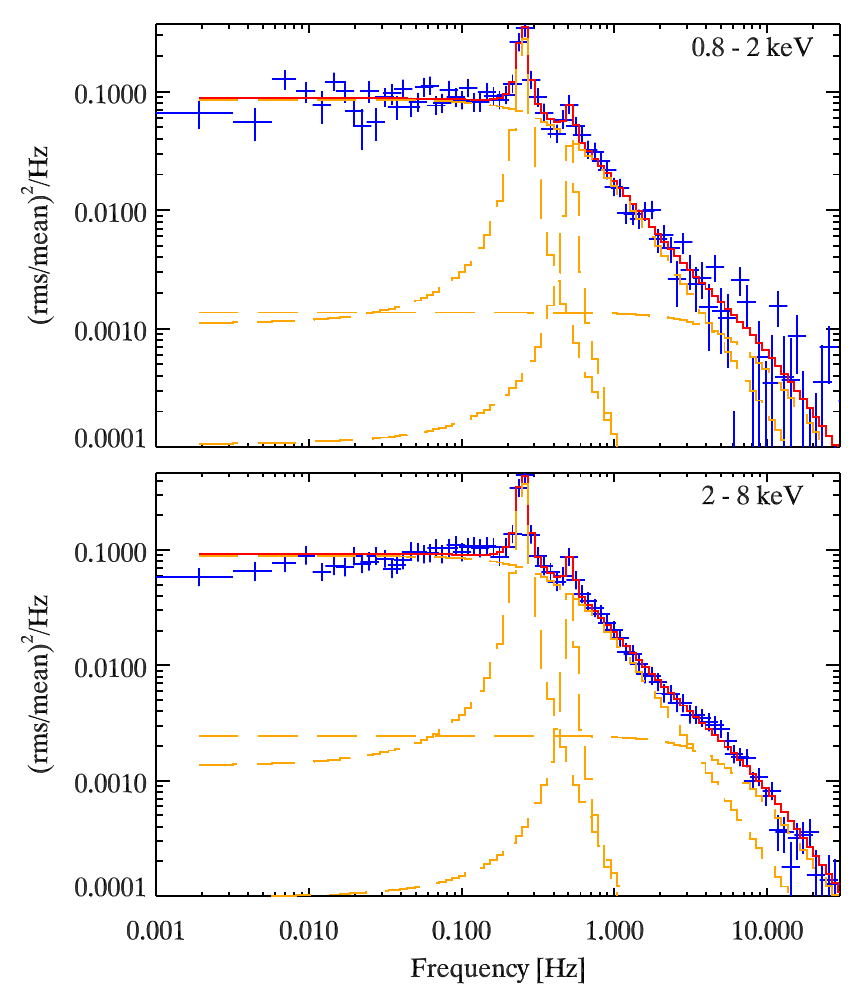}}
\caption{Power density specta obtained in the 0.8 -- 2 keV (upper panel) and 2 -- 8 keV (lower panel) energy ranges. The best fit model (red line) and the four individual components (orange dashed lines) are indicated.}
\label{Fig:pds}
\end{figure}

\begin{figure}
\resizebox{\hsize}{!}{\includegraphics[clip,angle=0]{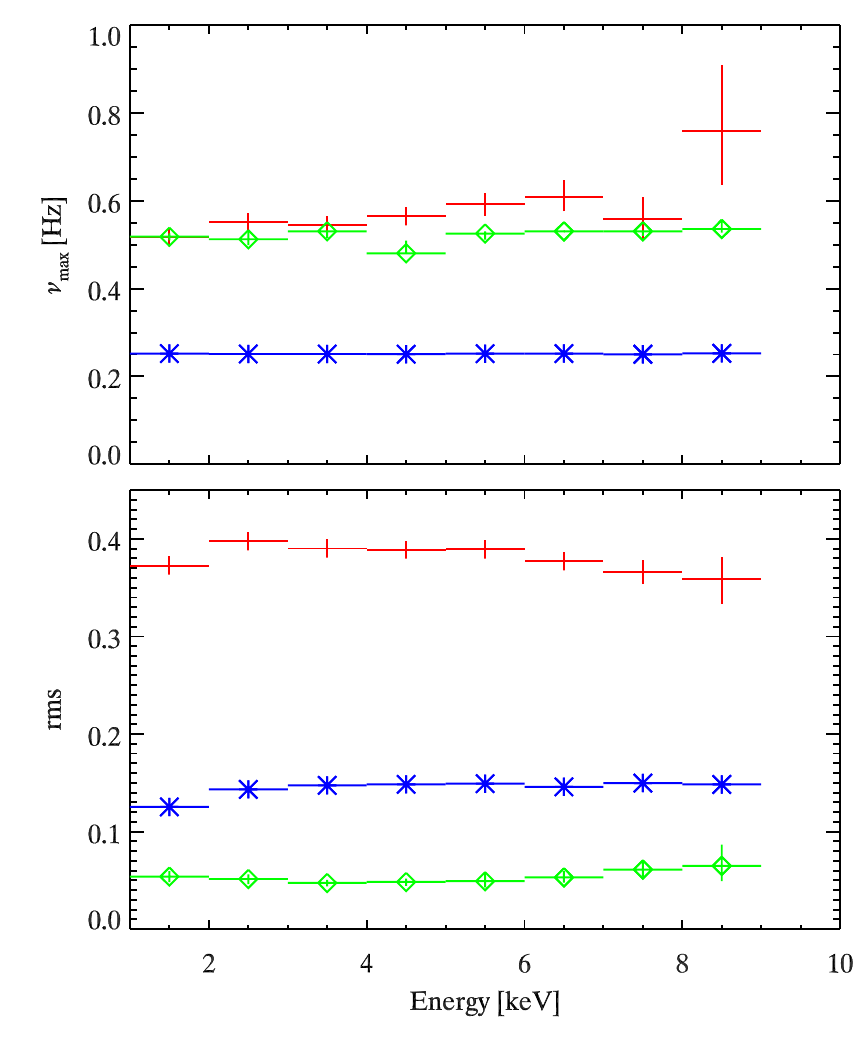}}
\caption{Energy dependence of the measured characteristic frequency (upper panel) and amplitude (lower panel). The BLN component is indicated by (red) crosses, the QPO and its harmonic are indicated by (blue) stars and (green) diamonds, respectively. The observed behaviour corresponds to the one normally seen during the LHS.}
\label{Fig:rms_spec_indv}
\end{figure}

\begin{figure}
\resizebox{\hsize}{!}{\includegraphics[clip,angle=0]{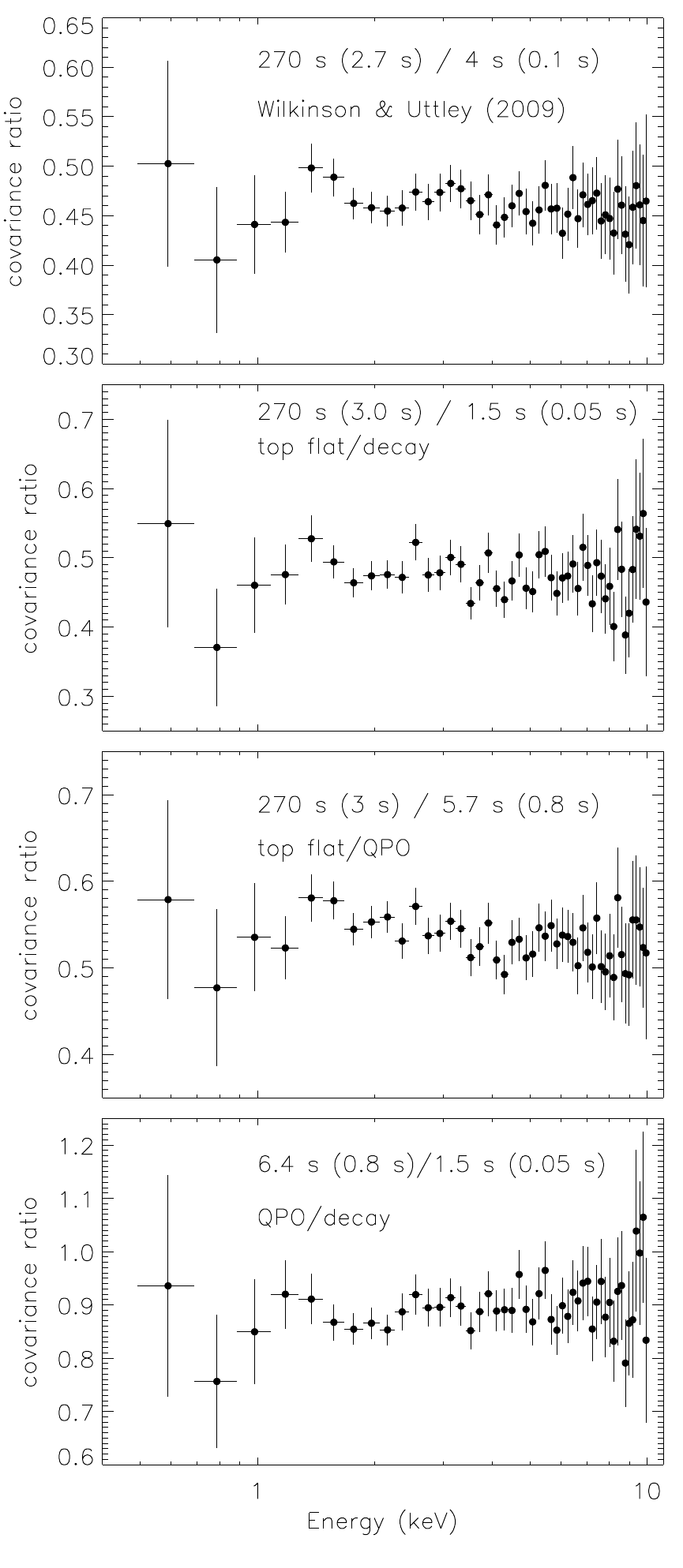}}
\caption{Covariance ratios derived by dividing covariance spectra on long timescales by those on short timescales. Time scales and binning used and corresponding components are indicated in each panel. In top panel \citet{2009MNRAS.397..666W} indicates that we used the time scales / binning of this paper.}
\label{Fig:covspecratio}
\end{figure}

\begin{table}
\caption{PDS parameters}
\begin{center}
\begin{tabular}{llll}
\hline\noalign{\smallskip}
 \multicolumn{1}{c}{component} & \multicolumn{1}{c}{rms [\%]} & \multicolumn{1}{c}{$\nu^{\mr{a}}$ [Hz]}  & \multicolumn{1}{c}{$\Delta^{\mr{b}}$ [Hz]}\\
 \hline\noalign{\smallskip}
 \multicolumn{4}{c}{0.8 -- 2 keV} \\
 \hline\noalign{\smallskip}
BLN 1 & $25.28^{+0.55}_{-0.73}$& 0$^{\mr{c}}$ &$0.48^{+0.02}_{-0.03}$\\
\noalign{\smallskip}
BLN 2 & $12.16^{+1.59}_{-1.65}$& 0$^{\mr{c}}$ &$6.9^{+3.1}_{-3.0}$\\
\noalign{\smallskip}
QPO & $12.69\pm0.32$& $0.2524^{+0.0009}_{-0.0008}$ & $0.0134^{+0.0013}_{-0.0012}$\\
\noalign{\smallskip}
QPO uh & $5.90^{+0.60}_{-0.57}$&$0.515\pm0.006$ & $0.025^{+0.009}_{-0.008}$\\
\hline\noalign{\smallskip} 
 \multicolumn{4}{c}{2 -- 8 keV} \\
 \hline\noalign{\smallskip}
BLN 1 & $25.82^{+0.21}_{-0.22}$& 0$^{\mr{c}}$ &$0.48\pm0.01$\\
\noalign{\smallskip}
BLN 2 & $14.96^{+0.34}_{-0.35}$& 0$^{\mr{c}}$ &$5.8\pm0.4$\\
\noalign{\smallskip}
QPO & $14.63\pm0.23$& $0.2513\pm0.0005$ & $0.0124\pm0.0007$\\
\noalign{\smallskip}
QPO uh & $6.12^{+0.27}_{-0.28}$&$0.514\pm0.003$ & $0.021\pm 0.004$\\
\hline\noalign{\smallskip} 
\end{tabular} 
\end{center}
\normalsize
Notes: \\
$^{\mr{a}}$: centroid frequency\\ 
$^{\mr{b}}$: half width at half maximum (HWHM)\\ 
$^{\mr{c}}$: zero centered Lorentzian
\label{Tab:bbpds}
\end{table}

 \begin{figure}
\resizebox{\hsize}{!}{\includegraphics[clip,angle=0]{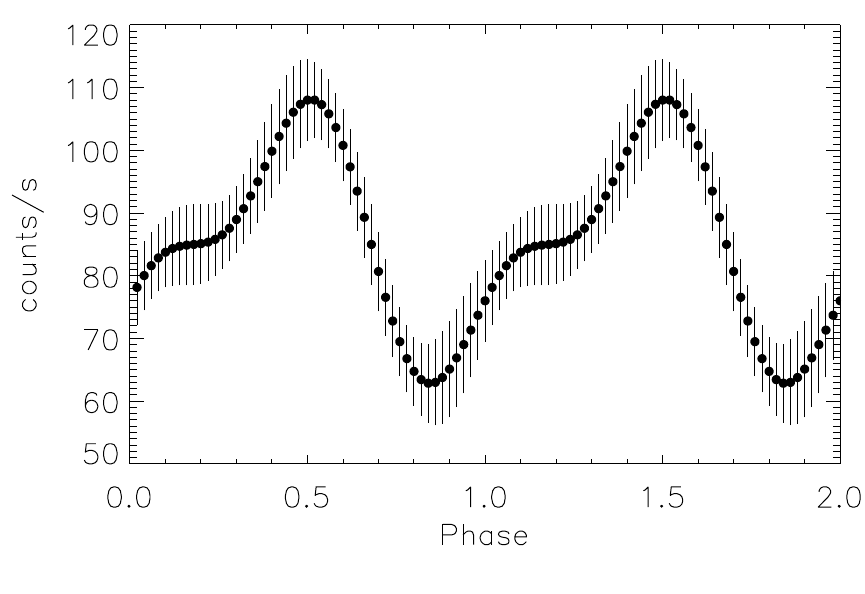}}
\caption{Reconstructed QPO waveform. For clarity, two cycles are plotted.}
\label{Fig:WF}
\end{figure}
 
\begin{figure}
\resizebox{\hsize}{!}{\includegraphics[clip,angle=0]{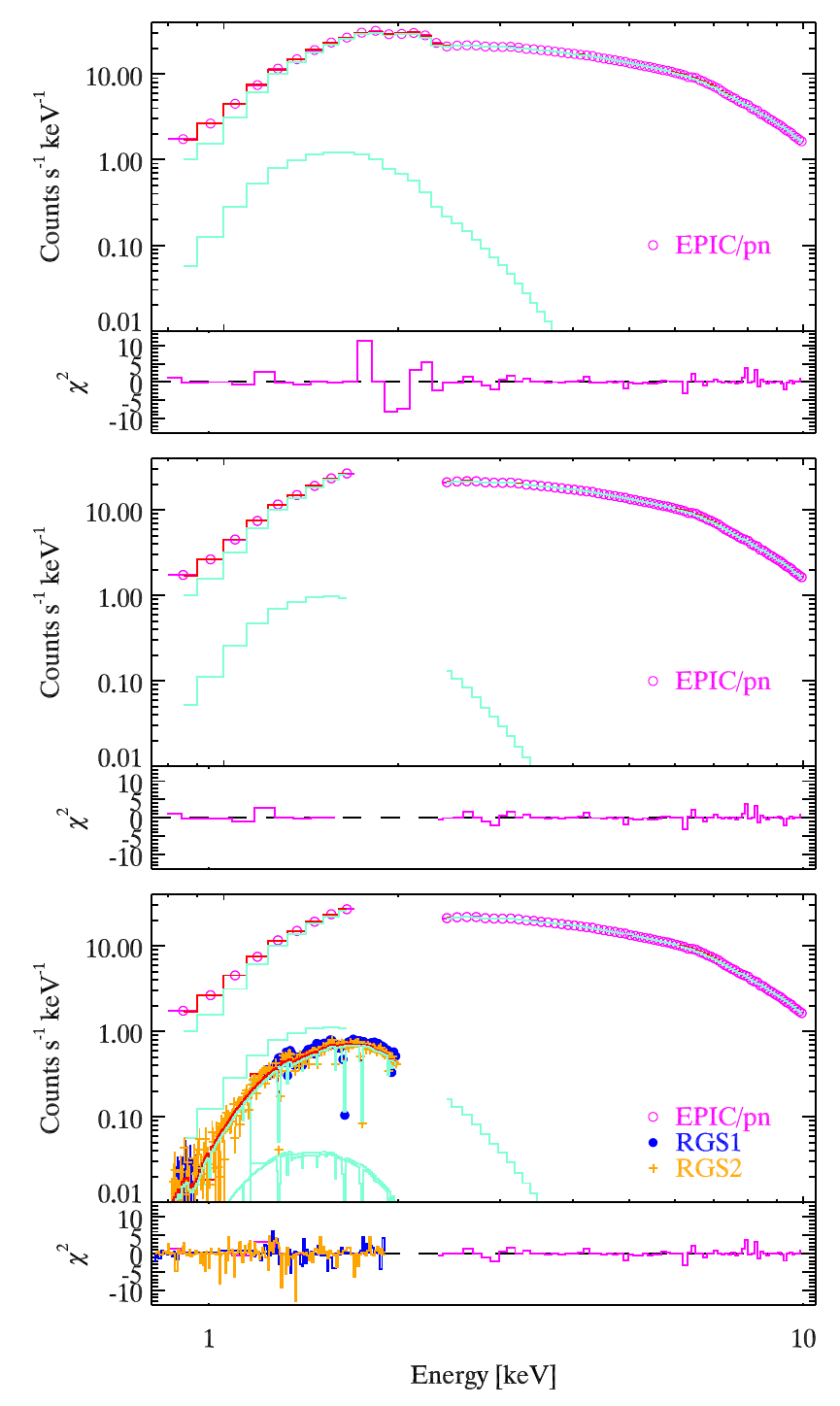}}
\caption{Combined RGS and EPIC/pn spectra in the 0.8 -- 2 keV and 0.8 -- 10 keV range. The best fit model and individual components are indicated. Details on the model best-fit can be found in Sec.~\ref{Sec:spec_prop} and Table~\ref{Tab:Spec_par}.}
\label{Fig:espec}
\end{figure}

\subsection{Timing properties}
The broadband PDS (2 -- 8 keV) can be well fitted by two zero centered Lorentzians describing the BLN components and two peaked Lorentzians for the QPO and its upper harmonic located at $0.2513\pm0.0005$ and $0.515\pm0.003$ Hz (Fig.~\ref{Fig:pds}). In the soft band (0.8 -- 2 keV) PDS the same components are present and we do not see any hint of additional disc variability at low frequencies. In the soft band the variability in the QPO and in BLN~2 are smaller than in the 2 -- 8 keV band. All parameters are given in Table~\ref{Tab:bbpds}. To study the energy dependence of the rms amplitude and of the characteristic frequency \citep[$\nu_{\mr{max}}=\sqrt{\nu^2+\Delta^2}$, where $\nu$ is the centroid frequency, and $\Delta$ is the half width at half maximum][]{2002ApJ...572..392B},  we fit PDS obtained in energy bands with a width of 1~keV. These narrow band PDS are well described by a single BLN component (BLN~1) and the QPO and its upper harmonic. Naturally, the narrow energy bands contain significantly less photons than the broad 2 -- 8 keV band which leads to a reduced signal-to-noise ratio in the PDS which becomes quite evident as significantly increased error bars at high frequencies (\ie\ short time scales) that hamper constraining BLN~2. This effect is already evident from a comparison of the soft and broadband PDS shown in Fig.~\ref{Fig:pds}. The rms spectrum of the BLN component (red crosses in lower panel of Fig.~\ref{Fig:rms_spec_indv}) shows a slight decrease with increasing energy. This kind of energy dependence is observed during the LHS only. In a study of XTE\,J1650--500 and XTE\,J1550--564 \citet{2005MNRAS.363.1349G} found that the rms spectrum in the LHS at energies below 10 keV is either flat or decreases with increasing energy, as seen here. The rms spectra of other states presented in \citet{2005MNRAS.363.1349G} do not show this kind of energy dependence at energies below 10 keV. A similar shape of the rms spectrum as found here has been observed for the BLN of the 2008 observation of \h1743\ -- where the BLN was observed at a similar frequency as here -- and for the BLN component with the highest characteristic frequency -- which was above 1 Hz -- in two observations of GX\,339--4 taken during LHS \citep{2015MNRAS.452.3666S}. 

The measured characteristic frequency of the BLN as well as of the QPOs does not show a strong energy dependence and stays rather flat (upper panel of Fig.~\ref{Fig:rms_spec_indv}).  According to our study on the BLN in a sample of BHTs \citep{2015MNRAS.452.3666S} this behaviour of the BLN is seen in the early phases of the LHS, while later on the characteristic frequency is lower at softer energies.

The value of the characteristic frequency of the QPO, the similar shape of the PDS at soft and harder energies, the shape of the rms spectrum, and the energy dependence of the characteristic frequency of the BLN are all consistent with \h1743\ being in the LHS at the time of our \xmm\ observation.

\subsubsection{Covariance ratios}
In addition, we derived covariance spectra on short and long timescales and obtained covariance ratios dividing the long timescale covariance spectrum by the short one. Deriving ratios of covariance spectra we can remove features in the spectra that are caused by gain shift due to Charge-transfer inefficiency and excess emission below $\sim$1.3 keV, as these effects affect the spectra in the same way on the time scales used in this paper. Thus, using ratios of spectra we can go to lower energies without worrying about spectral distortion due to the soft excess emission. Using the same timescales as in \citet{2009MNRAS.397..666W}: shorter time scales 0.1~s time bins measured in segments of 4~s; longer time scales 2.7~s time bins in segments of 270~s; the ratio (long/short) of these two spectra (shown in Fig.~\ref{Fig:covspecratio}) is mostly flat down towards the lower end of the energy scale contrary to the increase of the covariance ratio towards lower energies, which has been reported by \citet{2009MNRAS.397..666W} who studied \gx339\ and Swift\,J1753.5-0125 during a later phase of the LHS. Taking a look at the PDS (Fig.~\ref{Fig:pds}) we find that the short timescale 0.25 Hz start frequency bin is the characteristic QPO frequency, and the 5 Hz end bin is the frequency of BLN~2. We adjust the short timescale range using 0.05~s time bins measured in segments of 1.5~s. This range now comprises the decaying part of the PDS at higher frequencies, excluding the QPO and its upper harmonic. To make sure that the longer timescale range is also not affected by the QPO, we use 3~s time bins in segments of 270~s. Comparing covariance on long to short timescales we still obtain a flat ratio (see Fig.~\ref{Fig:covspecratio}). In addition, we define a time range that contains the QPO and its upper harmonic using 0.8~s time bins in segments of 6.4 or 5.7~s, to avoid overlap with the longer timescale range defined above. The covariance ratios derived from the frequency range of the QPO and of frequency ranges at higher or lower frequencies have also an overall flat shape (Fig.~\ref{Fig:covspecratio}). In the energy range between 1 -- 3 keV there may be some additional variability in the flat top noise compared to the QPO frequency range, while there seems to be additional  variability in the decaying part over the QPO frequency range around 2 keV. To exclude that the results shown here are affected by intrinsic time lags in the data, we measured cross-spectral phase-lags between various energy bands, and we have confirmed that the lags between hard and soft band variations are smaller than the time bin sizes used to make the long and short time-scale covariance spectra \citep{2009MNRAS.397..666W}.

\subsubsection{QPO waveform}
\label{Sec:QPOwf}
\citet{2015MNRAS.446.3516I} presented an approach to reconstruct the QPO waveform assuming a periodic function and using the amplitudes of and the phase difference between the QPO (or fundamental or first harmonic) and its upper (or second) harmonic.  
This reconstruction requires that a well-defined average underlying waveform exists, which can be tested by measuring that the ratio of the harmonic amplitudes and the phase difference between harmonics are clearly distributed around a mean value. In case of a highly non-linear decohering process a bias to the estimated underlying waveform may be introduced or it may be difficult to define a true waveform. As we do not have a full understanding of all the processes decohering the QPO, we define the waveform as a periodic function with the average QPO properties.

We split the light curve of 31.8~ks continuous exposure into $M=520$ segments, with each segment containing $N=2048$ time bins of a duration that equals five times the frame time. These values are chosen based on the assumption that the QPO can be regarded as periodic for $Q$ cycles, where $Q=\nu/(2\Delta)$ is the quality factor. As the broadband PDS (Fig.~\ref{Fig:pds}) shows a QPO and its upper harmonic at $0.2513\pm0.0005$ and $0.515\pm0.003$ Hz we can use the centroid frequency and width of these features to measure the relative strength of each harmonic and the phase difference between the harmonics in each segment. The technical details can be found in \citet{2015MNRAS.446.3516I}. The resulting averaged waveform is shown in Fig.~\ref{Fig:WF}. 

\subsection{Spectral properties}
\label{Sec:spec_prop}
We fitted the averaged EPIC/pn spectrum within \textsc{isis} V.~1.6.2 \citep{2000ASPC..216..591H} in the 0.8 -- 10 keV range, grouping the data to a minimum number of 20 channels and to a minimum signal-to-noise ratio of 3.  Grouping data allows us to use $\chi^2$ minimisation to obtain the best fit. We included a systematic uncertainty of 1 per cent. We used an absorbed (\texttt{tbabs}) disc blackbody (\texttt{diskbb})  plus a thermal Comptonisation component (\texttt{nthcomp}) with a Gaussian to attribute for emission of the Fe line at $\sim$6.7 keV. We included additional Gaussian components to model the excess emission below 1.3 keV and the features caused by gain shift due to Charge-transfer inefficiency around 1.8 and 2.2 keV \citep{2011MNRAS.411..137H,2014A&A...571A..76D}. This resulted in a fit with $\chi^2_{\mr{red}}=$ 63.2/84 (Fig.~\ref{Fig:espec}). The individual  spectral parameters are given in Table~\ref{Tab:Spec_par}. To check how the features around 1.8 and 2.2 keV affect the spectral parameters we excluded the energy range between 1.7 and 2.3 keV from the EPIC/pn spectrum and repeated the fit with the model used above ($\chi^2_{\mr{red}}=$38.4/76; Fig.~\ref{Fig:espec}). We found an increase in the inner disc radius and a slight decrease of the inner disc temperature (see Table~\ref{Tab:Spec_par}). Nevertheless, the values of both parameters are consistent within error bars in both cases. To investigate the effect of the soft excess emission on the parameters of the thermal component, we included RGS data using the 0.8 -- 2 keV range. As we are only interested in the general shape of the RGS spectra and not in individual absorption or emission lines in this energy range, we grouped RGS data to a minimum number of 5 channels and to a minimum signal-to-noise ratio of 3.\@ Using the same model as above we obtained a fit with a $\chi^2_{\mr{red}}$ of 389.7/333 (Fig.~\ref{Fig:espec}). The individual spectral parameters are consistent with the ones obtained from EPIC/pn data only (Table~\ref{Tab:Spec_par}). Including RGS data allowed us to constrain better the spectral parameters of the disc blackbody component. 

The photon index of $1.52^{+0.01}_{-0.02}$ is at the low end of the photon index range seen in the LHSs of \h1743\ \citep{2013MNRAS.429.2655S}, providing spectral evidence that the source was still too hard for the hard-to-soft state transition to occur. The low photon index is consistent with the photon indices of later \textit{Swift} observations that showed that the source stayed in the hard state during the entire outburst. 

From our spectral fits we obtained a foreground absorption of N$_{\mr{H}}=2.01^{+0.02}_{-0.01}\times 10^{22}$ atoms cm$^{-2}$. A similar foreground absorption has been obtained from three INTEGRAL, seven \textit{Chandra}, and three Suzaku observations \citep{2003A&A...411L.421P,2006ApJ...646..394M,2006ApJ...636..971C,2014ApJ...789..100S}. The INTEGRAL observations and four of the \textit{Chandra} observations were taken during outburst in 2003, while the remaining \textit{Chandra} data were obtained in 2004 when \h1743\ was in quiescence. The Suzaku observations were taken in 2012. A lower foreground absorption of N$_{\mr{H}}=1.6\pm0.1\times 10^{22}$ atoms cm$^{-2}$ has been reported based on a \textit{Swift}/XRT observation in Window-Timing mode obtained during the 2008 outburst \citep{2009MNRAS.398.1194C}. 

An inner disc radius of around 30 km, similar to the value obtained in our study, has been derived from the disc-blackbody normalization during the 2008 (failed) outburst using RXTE/PCA observations \citep{2010MNRAS.408.1796M}. The inner disc temperature from our \xmm\ observation is much lower than the temperatures reported from the 2008 outburst with RXTE \citep{2010MNRAS.408.1796M}. This can be explained by the fact that the PCA instrument covered a harder energy band and thus only energy spectra obtained during a later phase of the outburst  when the accretion disc is hotter require a thermal component. In the study of the 2008 outburst reported by \citet{2010MNRAS.408.1796M} spectral fits of most observations did not require a thermal component, and only for observations corresponding to the softest points in the HID of the failed outburst a disc blackbody component, with T$_\mr{in}\sim0.74$, was needed. From the combined spectral fit we obtain an unabsorbed flux $F_{\mr{0.8-10 keV}}=9.23$\ergcm{-9} in the 0.8 to 10 keV energy range, corresponding to a luminosity $L_{\mr{0.8-10 keV}}=7.97$\ergs{37}, assuming a distance of 8.5 kpc \citep{2012ApJ...745L...7S}.

\begin{figure}
\resizebox{\hsize}{!}{\includegraphics[clip,angle=0]{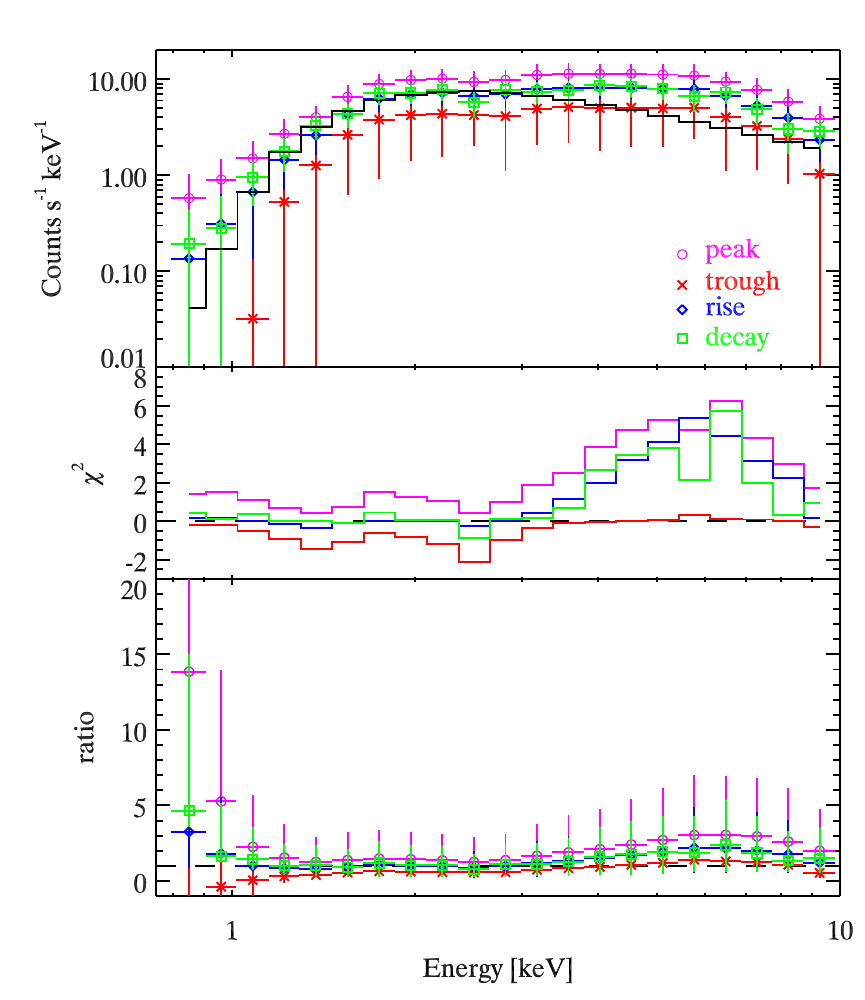}}
\caption{Spectra corresponding to four QPO phases. Phases are selected to be representative of rising ($\phi=0.3125$), peak ($\phi=0.5$), falling ($\phi=0.6875$) and trough ($\phi=0.875$) intervals. The line indicates the best-fit model obtained from the averaged spectrum. In addition, residuals between the phase resolved spectra and the best-fit model are shown.}
\label{Fig:QPOphspec}
\end{figure}

\subsubsection{QPO phase resolved energy spectra}
\label{Sec:qpo_spec_prop}
To obtain QPO phase resolved energy spectra, we can reconstruct the QPO waveform in each energy channel following Sec.~\ref{Sec:QPOwf}. Alternatively, we can make use of the fact that we have already reconstructed the full band QPO waveform and measure the phase lag at each harmonic between each energy channel and the full band \citep{2015MNRAS.446.3516I} . This approach allows us to maximise signal to noise. 

In Fig.~\ref{Fig:QPOphspec} we plot spectra for four QPO phases, along with the best-fit model obtained for the averaged spectrum. In addition, residuals between the phase resolved spectra and the best-fit model are shown. Comparing the phase resolved spectra to the best-fit model we see indications of spectral pivoting. Spectral pivoting with QPO phase is also observed in GRS\,1915+105 \citep{2015MNRAS.446.3516I}. For the spectrum corresponding to the trough interval the Comptonisation part agrees with the best-fit model, while the data point at lower energies, where the disc component contributes to the spectrum, are shifted to lower count rates compared to the model. For the remaining three phase resolved spectra there is better agreement of the disc dominated part of the spectra, while the Comptonised emission seems to be harder or containing additional contribution of a reflection component. Comparing the shape of the phase resolved spectra to the one of the best-fit model, the pivot takes place around 3 -- 4 keV.

\begin{table*}
\caption{Spectral parameters obtained by fitting the averaged RGS \& EPIC/pn spectra.}
\begin{center}
\begin{tabular}{llll}
\hline\noalign{\smallskip}
 \multicolumn{1}{c}{component} & \multicolumn{1}{c}{EPIC/pn full} &  \multicolumn{1}{c}{EPIC/pn gap} &  \multicolumn{1}{c}{RGS \& EPIC/pn gap} \\
 \hline\noalign{\smallskip}
N$_{\mr{H}}$ [cm$^{-2}$]& $2.01^{+0.04}_{-0.02}\times 10^{22}$ &$2.01^{+0.04}_{-0.06}\times 10^{22}$ & $2.01^{+0.02}_{-0.01}\times 10^{22}$\\
\noalign{\smallskip}
R$_{\mr{in}} [km]^{\ddagger}$& $29^{+16}_{-15}$ & $37^{+31}_{-28}$ & $35^{+20}_{-14}$\\
\noalign{\smallskip}
T$_{\mr{in}}$ [keV]& $0.34^{+0.05}_{-0.04}$& $0.30^{+0.10}_{-0.04}$& $0.30^{+0.03}_{-0.02}$\\
\noalign{\smallskip}
$\Gamma$ & $1.52^{+0.02}_{-0.06}$ &$1.52_{-0.05}^{+0.02}$& $1.52^{+0.01}_{-0.02}$\\
\noalign{\smallskip}
kT$_{\mr{e}}$ [keV]& $9^{+10}_{-2}$  & $9^{+7}_{-2}$& $9\pm1$\\
\noalign{\smallskip}
N$_{\mr{nthcomp}}$ & $0.163^{+0.005}_{-0.007}$ & $0.168^{+0.005}_{-0.013}$ & $0.168^{+0.003}_{-0.005}$ \\
\noalign{\smallskip}
N$_{\mr{Fe}}$ & $3.9^{+1.1}_{-0.9}\times10^{-4}$ & $3.8^{+1.1}_{-1.0}\times10^{-4}$ & $4.0^{+0.6}_{-0.5}\times10^{-4}$  \\
\noalign{\smallskip}
E$_{\mr{Fe}}$ [keV]& $6.73^{+0.10}_{-0.11}$ & $6.73\pm0.11$ & $6.73^{+0.06}_{-0.05}$ \\
\noalign{\smallskip}
$\sigma_{\mr{Fe}}$ [keV]& $0.47^{+0.15}_{-0.12}$ & $0.46^{+0.13}_{-0.11}$ & $0.46^{+0.07}_{-0.06}$ \\
\noalign{\smallskip}
C$_{\mr{rgs1}}$  & & & $1.054\pm0.007$ \\
\noalign{\smallskip}
C$_{\mr{rgs2}}$   & & & $0.999\pm0.007$ \\
\hline\noalign{\smallskip} 
\end{tabular} 
\end{center}
\normalsize
Notes: \\
$^{\ddagger}$: derived from the disc-blackbody normalization, assuming a distance of $8.5\pm0.8$ kpc and an inclination of $\theta = 75\pm3$\deg\ \citep{2012ApJ...745L...7S}\\
\label{Tab:Spec_par}
\end{table*}

\section[]{Discussion}
\label{Sec:dis}

From the evolution of the source during previous outbursts we expected to catch \h1743\ during the transition from the LHS to the HSS. The fact that neither the general shape of the observed PDS nor the characteristic frequency shows an energy dependence suggests that the source is actually observed during a state that is consistent with the LHS during an early rising phase of the outburst but at an X-ray luminosity high enough for the hard-to-soft transition to occur. 
The spectral parameters also indicate that the source is in the LHS at the time of our \xmm\ observation. The observed photon index is at the low end of the photon index range seen in the LHSs of \h1743\ \citep{2013MNRAS.429.2655S}, providing spectral evidence that the source was still too hard for the hard-to-soft state transition to occur. The \textit{Swift} follow-up observations show that \h1743\ remains in the LHS during its entire 2014 outburst (see Fig.~\ref{Fig:HID}). 

The QPO phase resolved spectra show indications of spectral pivoting. A similar result has been obtained for one of the observations studied in \citet{2015MNRAS.446.3516I}. Interestingly, \citet{2015MNRAS.446.3516I} observe a decrease of the ratio of the QPO phase resolved spectra to the mean spectrum with lower energy during QPO peak and an increase during trough, while we observe an increase with lower energy during QPO peak and a decrease during trough. 
Spectral pivoting has been reported for hard state observations of Cyg\,X--1 with a pivot energy of $\sim$50 keV \citep{2002ApJ...578..357Z}, so much higher than the 3 -- 4 keV range here. 
In this study the spectral pivoting has been addressed to changes of the accretion geometry that affect the seed photon population and lead to spectral pivoting on time scales of days. The spectral changes related to QPO phases suggest that type-C QPOs are caused by changes in Comptonization or in the reflection hump that take place on time scales of a few seconds. Spectral changes on such short time scales may be driven by changes in the accretion geometry. A model that relates type-C QPOs to a geometric origin is the Lense-Thirring precession model \citep{1998ApJ...492L..59S,2009MNRAS.397L.101I}. 

At the moment, there is no general agreement on the reason why BHTs sometimes tend to stay in the LHS and just start to decrease in luminosity after some time, while during other outbursts they go through the whole cycle including the soft states. In a systematic study of the hard-to-soft state transitions in Galactic X-ray binaries, \citet{2009ApJ...701.1940Y} find that the transition luminosity actually correlates with the rate-of-increase of the X-ray luminosity among outbursts or flares in persistent X-ray binaries. They suggest that the outbursts or flares during which the sources stay entirely in the hard state correspond to those cases with relatively higher rate-of-increase of the mass accretion rate, and the expected transition luminosity due to the correlation is substantially higher than the observed X-ray luminosity a source can reach. A previous comparative study of the timing and spectral properties of \h1743\ during its 2008 and 2009 outbursts finds that the rms values of the PDS and the centroid frequencies of the type-C QPOs are consistent between the two outbursts, as is the evolution of the spectral parameters \citep{2010MNRAS.408.1796M}. Investigating correlations between the centroid frequency of the QPO and spectral parameters \citet{2013MNRAS.429.2655S} finds that the correlations obtained from the 2008 ``failed'' outburst are consistent with correlations observed during other outbursts. 

With respect to the open question of an observable that allows to predict the subsequent outburst evolution it is interesting to notice that in the present observation the covariance ratio shows a slight decrease towards lower energies. In studies of later phases of the LHS of \gx339\ and Swift\,J1753.5-0125 \citet{2009MNRAS.397..666W} detect an increase in the covariance ratio towards lower energies that they interpret as additional disc variability on longer scales. In our recent studies of the BLN in black hole X-ray binaries \citep{2015MNRAS.452.3666S} we notice that for an observation of \h1743\ taken during early outburst rise (about four days before the onset of the RXTE monitoring) in 2008 the covariance ratio is flat. This implies that \h1743\ in the 2008 and 2014 outbursts lacks of coherent variability at soft energies below 1 keV observed by \xmm, indicating that the emission in the soft band is neither dominated by the Comptonized component nor the disc component, leading to a mixed and uncoupled variability. 

There are two possible explanations why covariance ratios of the two \h1743\ observations differ from those observed in other sources like \gx339\ or Swift\,J1753.5-0125. \h1743\ shows absorption dips \citep{2005ApJ...623..383H,2006ApJ...646..394M}, which indicates that the inclination of the source is relatively high, probably around 80\deg, while the remaining sources studied in \citet{2015MNRAS.452.3666S} have lower inclination \citep[below 70\deg; see appendix B of][for a summary on the inclination of 13 X-ray binaries]{2015MNRAS.447.2059M}. The fact that we see \h1743\ more edge-on than other sources like \gx339\ can explain why the contribution of additional disc emission on longer time scales does not show up in the covariance ratio of \h1743. Another possible explanation can be related to the fact that both observations of \h1743\ are taken during a failed outburst. In this case the presence or absence of additional soft band variability on longer time scales, which might indicate effective propagation of the mass accretion rate variation in the disc flow towards the emission region responsible for the soft band emission, might be the signature whether a source goes into a full or failed outburst, respectively. At the moment, this is speculative and further studies including more data on failed outbursts and on high-inclination sources would be needed to test this picture.     

\section*{Acknowledgments}
We would like to acknowledge useful discussions with Phil Uttley and Tomaso Belloni. We would like to thank Hui Zhang and Wenda Zhang for helping on requesting and promptly analysing Swift ToO observations of \h1743. This work was supported by the National Natural Science Foundation of China under grant No. 11333005, and 11350110498, by Strategic Priority Research Program "The Emergence of Cosmological Structures" under Grant No. XDB09000000 and the XTP project under Grant No. XDA04060604, by the Shanghai Astronomical Observatory Key Project and by the Chinese Academy of Sciences Fellowship for Young International Scientists Grant. This work makes use of software tools provided by Simon Vaughan. 

\bibliographystyle{mn2e}
\bibliography{}

\appendix

%\section[]{Online Material}
%\include{}
\bsp

\label{lastpage}

\end{document}